# Enhancement in Ground State Antihydrogen Formation via Three Body Recombination through Multiphoton effects

**S. Ghosh Deb and C. Sinha**

Department of Theoretical Physics,
Indian association for the Cultivation of Science,
Jadavpur, Kokata -700032, India.

e-mail:chand_sin@hotmail.com

**PACS. No**.: 34.80.Lx, 34.50.Rk and 36.10.-k

**Abstract**

  We study quantum mechanically for the first time, the antihydrogen ($\overline{H}$) formation cross sections in ground and excited (2s, 2p) states via the Three Body Recombination process inside a trapped plasma of antiproton and positron or in the collision between the two beams of them in presence of an external laser field, supposed to be the most efficient mechanism for cold and trapped antihydrogen. The laser polarization is chosen to be either parallel ($\parallel^l$) or perpendicular ($\perp^r$) to the incident $e^+$ momentum. The laser modification is found to be quite significant both qualitatively and quantitatively particularly for the ground state. We show that due to multiphoton effects, the ground state $\overline{H}$ formation is enhanced significantly while the excited states are somewhat suppressed w.r.t the corresponding field free (FF) results, in contrast to the single photon case where all the states (1s, 2s, 2p) are suppressed. These findings might have important implications for future $\overline{H}$ studies. The $\parallel^l$ polarization is found to be much more efficient than the $\perp^r$ one for the production of low energy $\overline{H}$. The $\perp^r$ geometry is quite sensitive with respect to the azimuthal angle of the outgoing spectator positron.

**Introduction**

Of late, one of the major challenges to the physicists (both experimental and theoretical) is to verify the fundamental symmetries of physics e.g., the CPT invariance and the gravitational – weak equivalence principle for antimatter, since any violation of the CPT symmetry would require a new physics beyond the Standard Model of the Particle Physics. Antihydrogen ($\overline{H}$) atom, being the most simple and stable than any other exotic atoms is an ideal antimatter for such verification by comparing its high precision ground state spectroscopy with the corresponding hydrogen spectroscopy that has already reached extremely high precision. In order to achieve the long term goal of high precision spectroscopy as well as of the antimatter gravitation studies, cold and trapped ground state $\overline{H}$ is badly needed in significant quantity.

For a significant production of cold and trapped $\overline{H}$, the most efficient mechanism is the Three Body Recombination (TBR) in a plasma of antiprotons ($\overline{p}$) and positrons ($e^+$), as compared to other $\overline{H}$ formation processes [1-41], e.g. the radiative recombination (RR) [31-33, 41], the three body charge transfer (CT) between the Ps and $\overline{p}$ [34-40]. One of the main reasons for this is that in the TBR the spectator $e^+$ efficiently carries off the extra energy unlike the RR process where the time required to radiate the photon is typically longer than the duration of collision between the $\overline{p}$ and the $e^+$ [21]. The main disadvantage of the TBR is that the $\overline{H}$ is favorably formed in the excited states, while in the RR process the ground state is favored [20]. However, in the latter case (RR) the cross section itself is much lower particularly at low incident energies [32], although the RR dominates the TBR at higher energies [30]. Further the LA RR cross sections were found (theoretically) to be even lower than the field free (FF) [30]. However an enhancement was observed experimentally [42] for laser stimulated RR process using beams of electron, proton and laser photon to produce $\overline{H}$ atom.

The recent experiments [1-20] at CERN have been able to produce cold ( ~ meV ) $\bar{H}$ using a nested Penning trap via the TBR process, which normally favors its formation in highly excited states and then through cascading they could finally reach to the ground state.

Theoretically, Robicheaux [23] performed the first simulations ( using CTMC ) followed by a series of works [24 – 28] attempting to incorporate the experimental conditions [1 -20] for the production of $\bar{H}$ using the same TBR process. However, it should be mentioned that the TBR rate for an ion introduced into magnetically confined, weakly correlated and cryogenic pure electron plasma was first calculated by Glinsky and O' Neil [22]. The Authors also predicted that the antimatter analogue of the electron-ion TBR process is a possible way of producing $\bar{H}$ for use in gravitational and spectroscopic studies [22].

Since nowadays in designing most of the collisional experiments, an external field is involved, it should be quite worthwhile to study the above TBR process for $\bar{H}$ formation in the presence of an external laser field.

The present work addresses to the laser assisted antihydrogen formation to ground and excited states via the TBR process in a collision between an antiproton and two positrons, the active and the spectator one, required for the momentum and energy conservation in such process. To our knowledge, the present work is the first quantum mechanical attempt for the laser assisted TBR process.

As for the experimental situation, the first laser-controlled $\bar{H}$ production experiment was devised by the ATRAP collaboration [5, 6] that might give colder atoms, although their temperature has not yet been measured, the experiment being still at its infancy. Later, ATHENA [8] studied experimentally the laser-induced $\bar{H}$ formation in the highly excited states ( n=11 ), where no laser effect was observed on the $\bar{H}$ formation. The new generation experiment, an international project named as ALPHA [12-16] studied the $\bar{H}$ production using a lower magnetic field (IT) than was used earlier [7]. However, to our knowledge, no laser-induced $\bar{H}$ formation study has yet been made with this device. In fact the primary goal of ALPHA [12] is to achieve the stable trapping of the $\bar{H}$ atoms to facilitate measurements of its properties. Very recently ALPHA [15] has demonstrated the first $\bar{H}$ formation in an octupole-based magnetic minimum neutral atom trap.

. The present model has certain limitations and assumptions in the context of the existing extreme low energy (~ meV ) experiments and was described in detail in our earlier work [30] for the FF situation., the energy range considered in the present work being relatively much higher (~5-50 eV) .

It may be pointed out here that in the present calculation we have considered single particle collision for the TBR reaction and as such we have not incorporated the density of the constituent particles. However it is implied that in order to dominate the TBR process over the RR one at lower temperatures, the density of the $e^+$ plasma should be reasonably high [20]. In most of the theoretical TBR calculations [22-28], the reaction (collision) rates are considered while the present study concentrates on the TBR cross sections only.

The collision dynamics (TBR) is treated quantum mechanically in the frame work of Coulomb modified eikonal approximation (CMEA) [30, 46] while the laser field is treated classically in the dipole approximation using the Coulomb gauge. The laser field dresses the two incoming positrons (active and the spectator) as well as the outgoing $e^+$ states non perturbatively by using eikonal modified Volkov wave functions while the dressing of the final bound state of the $\overline{\text{H}}$ atom (ground or excited) is treated perturbatively by solving the time dependent Schrödinger equation. The laser field is chosen to be much less than the atomic units of field strength ($5 \times 10^{11}$ V/m) in order to avoid the effect of field ionization and the frequency of the laser field is kept much below the binding energy of $\overline{\text{H}}$ (soft photon limit). Both the differential and the total $\overline{\text{H}}$ formation cross sections are studied for single as well as multiphoton exchange.

**Theory**

The present work deals with the following laser assisted (LA) three body recombination process (TBR) in the ground and excited states:

$$\overline{p} + e^+ + e^+ + N\gamma(\omega, \vec{\varepsilon}) \rightarrow \overline{\text{H}}(\overline{\text{H}}^*) + e^+ \qquad (1)$$

where N stands for the multiphoton absorption and emission and $\gamma(\omega, \vec{\varepsilon})$ denotes the laser photon with angular frequency $\omega$ and field strength $\vec{\varepsilon}_0$, the laser being chosen as continuum wave (CW). The first incoming positron is the active one to form the antihydrogen atom in the final state while the second positron acts as a spectator that carries away the excess energy and momentum released in the recombination process.

The prior form of the transition matrix element for the aforesaid TBR process (1) is given by [30] :

$$T_{if} = -i \int_{-\infty}^{\infty} dt \langle \Psi_f^-(\vec{r}_1, \vec{r}_2)(1+\vec{P}) | V_i | \psi_i(\vec{r}_1, \vec{r}_2) \rangle \qquad (2)$$

where $\vec{P}$ denotes the exchange operator corresponding to the interchange of the two positrons in the final channel and $V_i$ is the initial channel perturbation which is the part of the total interaction not diagonalized in the initial state and $\psi_i$ is the initial channel asymptotic wave function. We have used atomic unit ( a.u. ) through out the calculation.

The energy conservation relation for the TBR process is given by,

$$\frac{k_1^2}{2} + \frac{k_2^2}{2} \pm N\omega = \frac{k_f^2}{2} + \varepsilon_{\overline{H}} \qquad (3)$$

where $\vec{k}_1$ and $\vec{k}_2$ are the incident momentum of the active and the spectator positrons $(e^+ s)$ respectively; $\vec{k}_f$ being the final momentum of the outgoing spectator $e^+$, $\varepsilon_{\overline{H}}$ is the binding energy of $\overline{H}$. '+N' refers to absorption and '−N' refers to the emission of photons.

In the present model, the two $e^+$s are initially assumed to move parallel to each other, for the sake of simplicity. This assumption corresponds to a special case in the plasma environment, although it could probably be realized for a parallel beam of $e^+$s moving towards a trapped $\overline{p}$ in future experiments.

The total Hamiltonian (H) of the system for the reaction (1) may be written as

$$H = -\frac{1}{2}(i\vec{\nabla}_1 + \vec{A})^2 - \frac{1}{2}(i\vec{\nabla}_2 + \vec{A})^2 - \frac{1}{r_1} - \frac{1}{r_2} + \frac{1}{r_{12}} \qquad (4)$$

where $\vec{r}_1$ and $\vec{r}_2$ represent the position vectors of the active $e^+(\vec{r}_1)$ and the spectator $e^+(\vec{r}_2)$. The laser (CW) field is chosen to be a single mode linearly polarized, spatially homogeneous electric field represented by $\vec{\varepsilon}(t) = \vec{\varepsilon}_0 \sin(\omega t + \xi)$; $\xi$ being the initial phase of the laser field; the corresponding vector potential in the Coulomb gauge being $\vec{A}(t) = \vec{A}_0 \cos \omega t$ with $\vec{A}_o = \vec{\varepsilon}_0 / \omega$; $\xi$ is chosen to be zero in the present work.

The initial channel asymptotic wave function $\psi_i$ in equation (2) satisfies the following Schrödinger equation:

$$( -\frac{1}{2}(i\vec{\nabla}_1 + \vec{A})^2 - \frac{1}{2}(i\vec{\nabla}_2 + \vec{A})^2 - \frac{1}{r_1} - \frac{1}{r_2} - E ) \psi_i = 0 \tag{5}$$

and is given by,

$$\psi_i = \chi_{k_1} \chi_{k_2} \tag{6}$$

where $\chi_{k_i}$; $i=1, 2$ refers to the Coulomb Volkov (CV) solution [43, 44] and is given by,

$$\chi_{k_i} = N_i e^{i(\vec{k}_i \cdot \vec{r}_i + \vec{k}_i \cdot \vec{\alpha}_o \sin \omega t - E_{k_i} t)} {}_1F_1(i\eta_i, 1, -i(k_i r_i - \vec{k}_i \cdot \vec{r}_i)) \tag{7}$$

with $N_i = \exp(-\pi\eta_i / 2)\Gamma(1 - i\eta_i)$; $\eta_i = -\frac{1}{k_i}$; $\vec{\alpha}_0 = \vec{\varepsilon}_0 / \omega^2$. $\tag{8}$

and $\vec{k}_i$ denotes the momentum of the active $(\vec{k}_1)$ or the spectator $e^+$ $(\vec{k}_2)$.

The final channel wave function $\Psi_f^-$ satisfies the three-body Schrödinger equation obeying the incoming wave boundary condition:

$$(H - E)\Psi_f^- = 0 \tag{9}$$

where the dressed final channel wave function $\Psi_f^-$ is approximated in the framework of eikonal approximation as follows:

$$\Psi_f^- = \chi_{k_f}(\vec{r}_2, t)\phi_H^d(\vec{r}_1, t) \tag{10}$$

where $\chi_{k_f}(\vec{r}_2,t)$ represents the eikonal modified Volkov state of the outgoing scattered positron and is given by

$$\chi_{k_f} = \exp i(\vec{k}_f.\vec{r}_2 + \vec{k}_f.\vec{\alpha}_0 \sin \omega t - E_{k_f} t) \times \exp[i\eta_f \int_z^\infty (\frac{1}{r_{12}} - \frac{1}{r_2})dz'] \qquad (11)$$

with $\vec{\alpha}_0 = \vec{\varepsilon}_0/\omega^2$; $\eta_f = \dfrac{1}{\left|\vec{k}_f - \vec{A}(t)\right|}$ .

The distortions of the incident positrons are considered in both the channels through the Coulomb as well as the eikonal wave functions. The Coulomb modifications takes account mainly of the inner scattering region ( some account is taken for the outer region as well ) while the asymptotic region is mainly covered by the eikonal part. Thus the present approximation can be pushed to much lower energy regime than the plain eikonal approximation. However, the present theory is not supposed to be valid for the extreme low energy (~ meV) region at which the recent experiments [ 1 – 20 ] are performed.

The dressed ground state wave function of the $\overline{H}$ in the final channel is constructed using the first order perturbation theory in the Coulomb gauge and is given as:

$$\phi_{\overline{H}}^d(\vec{r}_1,t) = \frac{1}{\sqrt{\pi}} e^{-iW_0^{\overline{H}}t} e^{-\lambda_f r_1}[1 + i\vec{A}(t).\vec{r}_1] \qquad (12)$$

where $W_0^{\overline{H}}$ is the energy of the ground state $\overline{H}$; $\lambda_f$ being the bound state parameter of the $\overline{H}$ in the ground state (1s).

In presence of the laser field, the angular momentum $l$ is no more a good quantum number and as such the excited states (2s, 2p) lose their identity in presence of the field. In fact, since the dipole operator has a non-vanishing matrix element between the 2s and $2p_0$ states, these two states are mixed by the dipole perturbation and as such the dressed $\overline{H}$ formed in the excited state (2s or 2p) could be expressed as a superposition of the 2s and 2p$_0$ state (linear combination) as follows:

$$\phi_{\overline{H}}^d(\vec{r}_1,t) = \frac{1}{\sqrt{2}}(\psi_1(r_1)e^{-i/\hbar(E_{n=2}-\Delta E)t} \pm \psi_2(r_1)e^{-i/\hbar(E_{n=2}+\Delta E)t}) \qquad (13)$$

with

$$\psi_1 = \frac{1}{\sqrt{2}}(\psi_{200} + \psi_{210})$$

$$\psi_2 = \frac{1}{\sqrt{2}}(\psi_{200} - \psi_{210}) \; ; \qquad (14)$$

$\Delta E = 3\varepsilon/Z$ denotes the Stark shift in a.u., Z being the charge of the target atom. In equation (14) $\psi_1$ represents the lower energy state while $\psi_2$ represents the upper energy state.

We designate hence forth the lower energy state as 2s and the upper energy state as 2p. Since the laser field is always chosen along the direction of the polar axis, only m=0 state contributes to this LA process due to the dipole selection rule $\Delta$m=0.

Finally, after performing the time integration [45] in equation (2), the transition matrix element $T_{if}$ reduces to,

$$T_{if} = -i\frac{1}{(2\pi)^{1/2}}\sum_N \delta(E_{k_f} - E_{k_1} - E_{k_2} + N\omega)J_N(\vec{K}.\vec{\alpha}_0)I \qquad (15)$$

where $J_N$ is the Bessel function of order N; $\vec{K} = \vec{k}_1 + \vec{k}_2 - \vec{k}_f$; the integral I being the space part of the transition matrix element. It should be noted that in performing the time integration analytically, we approximate ( for the sake of simplicity ) the quantity $\vec{A}(t)$ in the eikonal phase term in equation (11) by its $t=0$ value $A_0$. We have checked that the eikonal phase factor is not very sensitive with respect to the time dependence of the vector potential and as such this is not supposed to be a very crude approximation.

The laser assisted differential cross section ( $\frac{d\sigma}{d\Omega}$ ) for the formation of $\bar{H}$ for N photon transfer is given as:

$$( \frac{d\sigma}{d\Omega} )_N = \frac{k_f}{k_1 k_2}[T_{if}]^2 = \frac{k_f}{k_1 k_2}[\frac{1}{4}| f + g |^2 + \frac{3}{4}| f - g |^2] \quad ; \quad (16)$$

f and g being the direct and exchange amplitudes respectively.

The total cross section (TCS) for a given value of N can be obtained by integrating the differential cross section equation (16) over the solid angle

$$\sigma_N = \int (\frac{d\sigma}{d\Omega})_N d\Omega \quad ; \quad (17)$$

and the total multiphoton cross section is given by

$$\sigma = \sum_N \sigma_N \quad (18)$$

We have calculated the cross sections for two directions of the laser polarization vector ($\vec{\varepsilon}_0$), e.g., $\vec{\varepsilon}_0$ is parallel to the incident positron's momentum $\vec{k}_i$ ($\vec{\varepsilon}_0 \parallel \vec{k}_i$, parallel geometry ) and $\vec{\varepsilon}_0$ is perpendicular to $\vec{k}_i$ ( $\vec{\varepsilon}_0 \perp \vec{k}_i$, perpendicular geometry ); $\vec{\varepsilon}_0$ being always along the polar axis. Some preliminary results for this process have been reported recently [46] only for the $\parallel^l$ geometry.

For $\vec{\varepsilon}_0 \parallel \vec{k}_i$, both $\vec{\varepsilon}_0$ and $\vec{k}_i$ are along the polar axis Z, while for $\vec{\varepsilon}_0 \perp \vec{k}_i$, since $\vec{\varepsilon}_0$ is chosen to be the polar axis, $\vec{k}_i$ is $\perp^r$ to the Z axis. For the $\parallel^l$ geometry the differential cross section (DCS) is merely $\theta_f$ (scattering angle) dependent while for the $\perp^r$ geometry the cross section depends both on $\theta_f$ and $\phi$ (azimuthal angle) of the outgoing spectator positron. The DCS results are, as a rule, presented in terms of the scattering angle (i.e. the angle between $\vec{k}_i$ and $\vec{k}_f$) and, conventionally, $\vec{k}_i$ or $\vec{q}$ ($\vec{k}_i - \vec{k}_f$) is chosen to be the polar axis. However, since the present theory is based on the choice that the polar axis is along the field direction $\vec{\varepsilon}_0$, the analysis requires a transformation of the coordinate system so that, in the final coordinate system, $\vec{k}_i$ becomes the z axis (new coordinate system). To do this the coordinate system is rotated about the Y axis in the counter clockwise direction through an angle $\chi$ ($\chi = 90^0$). The corresponding transformation matrix in terms of Euler angles [47] is given as,

$$\begin{bmatrix} x \\ y \\ z \end{bmatrix} = \begin{pmatrix} \cos\chi & 0 & -\sin\chi \\ 0 & 1 & 0 \\ \sin\chi & 0 & \cos\chi \end{pmatrix} \begin{bmatrix} X \\ Y \\ Z \end{bmatrix} \qquad (19)$$

where X, Y, Z refer to the old coordinate system while x, y, z refer to the new coordinate system.

**Results and discussions:**

The multiphoton differential (MDCS) as well as the total (MTCS) $\overline{H}$ formation cross sections in its ground and excited (n=2) states are computed via the TBR reaction (1) in the framework of CMEA. The laser field strength is chosen as $\varepsilon_0 = 5\times 10^9$ V/m and the laser photon energy $\hbar\omega = 1.17$ eV ( wavelength $\lambda = 1060$ nm ) corresponding to the Nd:Yag laser.

Figures 1 represent the multiphoton differential cross sections (MDCS) in the ground and excited states for some unequal $(E_1 \neq E_2)$ energies of the two incident positrons ($e^+$s), the active ($E_1$) and the spectator ($E_2$) for both the parallel ($\|^l$) and the perpendicular ($\perp^r$) geometries.

Comparing the partial MDCS curves in figure 1, it may be inferred that at low incident energies ( e.g., $E_1 = 10$ eV, $E_2 = 7$ eV ), the LA $\overline{H}$ cross sections follow the order 2s > 2p> 1s ( except for extreme forward angles) for the $\|^l$ geometry while for the $\perp^r$, the order is 2p>2s>1s through out the angular region. At intermediate (~15eV) and high incident energies, the order changes to $1s > 2s > 2p$ ( not shown ). Regarding the behavior w.r.t. to the energies, the field free (FF) nature is maintained qualitatively [30], i.e., for a fixed sum of $E_1$ and $E_2$, the partial MDCS follow the order $(E_1 > E_2) > (E_1 = E_2) > (E_1 < E_2)$ for both the geometries ( not shown ).

Quantitatively, the 1s MDCS for the $\parallel^l$ geometry dominates ( ~ by a factor of 2) over the $\perp^r$ one (vide fig. 1) in contrast to the excited states (2s, 2p), where the $\perp^r$ cross sections are slightly higher than the $\parallel^l$ one.

Next we come over to the multiphoton total cross sections (MTCS) for the $\overline{H}$ formation in figures 2 – 7 for different sets of incident energies using both the polarizations of the laser field.

Figures 2(a-c) display the partial MTCS against the active $e^+$ energy ($E_1$) for a fixed value of $E_2$ ( 7 eV ) using both the geometries along with the FF results. In the $\parallel^l$ geometry, the ground state MTCS enhances remarkably ( by order of magnitude ) with respect to FF (vide figure 2a) through out the energy range (0-50eV) while, in the $\perp^r$ geometry, the laser modification is comparatively less in the low energy region (0-15 eV) but for higher energies (>15 eV), the enhancement is quite significant (figure 2a). On the contrary, for the excited states (figures 2b & 2c), the laser field suppresses the FF cross sections almost through out the energy range (with some exceptions at extreme low energies for the 2s state) for both the geometries, the suppression being much more for the 2p state.

For lower $E_2$ ( fig.2 ), the $\parallel^l$ MTCS follows the same order 2s>2p>1s as noted for the MDCS ( fig. 1) at low $E_1$ while the $\perp^r$ MTCS maintains the order of the FF [30] i.e., 2p>2s>1s (figure 2). This indicates that the $\parallel^l$ polarization favors the meta stable 2s state than the dipole allowed 2p state (at low energy regime) while for the $\perp^r$ case the 2p state is preferred. This effect could probably arises due to the different orientations of the dipole moment of the dressed states $\psi_1$ and $\psi_2$ in Eq. (14), e.g., $\psi_1(\psi_2)$ being parallel ( anti parallel ) to the laser field ($\vec{\varepsilon}_0$) resulting in different strength of interactions of the outgoing $e^+$ with the $\overline{H}$ atom. In contrast, at higher $E_1$ ($\geq$ 15 eV), the 1s state is highly enhanced as compared to the 2s and 2p states ( for both the geometries) and the order is 1s>2s>2p for $\parallel^l$ polarization and 1s>2p>2s for $\perp^r$ polarization, which is quite different from the corresponding FF behavior [30] where the trend was 2s>2p>1s. This might be due to the fact that the polarization effect decreases with increasing energy.

On the other hand, at higher $E_2$, the excited state (2s, 2p) MTCS still dominates (over 1s) at extreme low energies as in the FF (vide inset fig. 3a) but the span of dominance ($E_1$) is reduced (~ 0-5 eV) as compared to FF (0-15 eV).

Figures 4(a-c) illustrate the individual photon distribution for the partial (1s, 2s, 2p) TCS for a fixed kinematics ($E_1 = 10\,\text{eV}$, $E_2 = 7\,\text{eV}$). Significant differences are noted in the photon distributions between the two geometries particularly for the 1s state as follows. The $\parallel^l$ spectrum is sharply peaked around N=0 while the $\perp^r$ spectrum is widely spread among the multiphoton processes. Figs. 4 also reveal that the emission processes are always much dominant than the absorption ones (for all the three states) as is expected for an exothermic reaction. Further, the $\parallel^l$ cross section is much more enhanced compared to the $\perp^r$ one. Comparing figures 2 (for MTCS) with the single photon TCS (figs. 4) it can be inferred that the latter are significantly suppressed w.r.t. the FF indicating the importance of the multiphoton effects in the TBR reaction.

Figure 5 describing the dependence of the MTCS on the azimuthal angle $\phi$ for the $\perp^r$ geometry ( the $\parallel^l$ geometry being independent of $\phi$), indicates that the 1s MTCS is more favoured for higher values of $\phi$ (e.g., $60^0$, $300^0$) than for $\phi = 0^0$. Further, the MTCS is found to be minimum for $\phi = 180^0$ which is expected physically due to strong repulsion between the two positrons. For the excited states, the qualitative behaviour is similar to the 1s state, although the value of $\phi$ (for the maximum MTCS) is different in each case. It is thus possible to control the $\overline{H}$ formation cross sections by varying the laser parameter $\phi$ for the $\perp^r$ geometry.

Figure 6, describing the MTCS versus the laser photon energy ($\omega$), demonstrates that for the $\parallel^l$ geometry, the curve has a maximum at the frequency $\omega \approx 0.8\,\text{eV}$, while in the $\perp^r$ case, instead of a distinct maximum only a valley is noted. An important distinction between the two geometries is that the $\parallel^l$ MTCS decreases gradually while the $\perp^r$ MTCS becomes almost insensitive with increasing $\omega$.

Finally, figure 7 exhibits the variation of the MTCS against the laser field strength $\varepsilon_0$ in atomic units (i.e. in units of $5 \times 10^{11}$ V/m). The MTCS is found to increase linearly with respect to $\varepsilon_0$ for the parallel geometry while for the perpendicular, the MTCS attains a parabolic nature, the enhancement being much greater for the former ($\parallel^l$) than for the latter ($\perp^r$).

**Table 1**: Power laws obeyed by the parallel partial MTCS for the ground (1s) and excited states (2s & 2p) for different incident energy ranges.

| Energy range (in eV) | Power law obeyed FF/ LA | | | |
|---|---|---|---|---|
| $E_1 = E_2$ | 1s | 2s | 2p | 1s+2s+2p |
| 5-10 | $E^{-2.9}/E^{-2.9}$ | $E^{-3.7}/E^{-2.1}$ | $E^{-4.8}/E^{-4.0}$ | $E^{-4.5}/E^{-4.15}$ |
| 10-25 | $E^{-3.9}/E^{-3.7}$ | $E^{-5.3}/E^{-5.3}$ | $E^{-6.0}/E^{-5.1}$ | $E^{-5.4}/E^{-4.6}$ |
| 25-50 | $E^{-5.0}/E^{-4.9}$ | $E^{-6.1}/E^{-6.9}$ | $E^{-6.7}/E^{-6.9}$ | $E^{-5.9}/E^{-5.2}$ |
| $E_1 \leq E_2 (= 7\,\text{eV})$ | | | | |
| 5-10 | $E_1^{-1.6}/E_1^{-1.04}$ | $E_1^{-1.3}/E_1^{-1.4}$ | $E_1^{-1.8}/E_1^{-1.3}$ | $E_1^{-1.6}/E_1^{-1.4}$ |
| $E_1 \geq E_2 (= 7\,\text{eV})$ | | | | |
| 10-25 | $E_1^{-1.5}/E_1^{-0.95}$ | $E_1^{-1.7}/E_1^{-1.5}$ | $E_1^{-2.1}/E_1^{-1.5}$ | $E_1^{-1.9}/E_1^{-1.3}$ |
| 25-50 | $E_1^{-1.5}/E_1^{-0.56}$ | $E_1^{-1.8}/E_1^{-1.9}$ | $E_1^{-2.3}/E_1^{-1.8}$ | $E_1^{-1.9}/E_1^{-1.1}$ |

Table 1 depicts that the energy (i.e. temperature) dependence of the partial MTCS does not follow any simple power law which corroborates (qualitatively) the experimental findings [4] as well as the CTMC calculation [27]. However, the present MTCS (1s+2s+2p) more or less follows the quasi-static prediction ($\sim T^{-9/2}$) in the FF situation [22] for the $E_1 = E_2$ case only. In contrast, for $E_1 \neq E_2$, the present power law is far away (lower) from the quasi-static behavior both for the LA and the FF cases. The most important findings from the table is that the LA power law falls off much slowly as compared to the FF, particularly for the ground state.

**Conclusion**

The salient features of the present study are as follows:

The ground state multiphoton cross sections (MTCS) are enhanced by order of magnitude with respect to the FF [26] (particularly for the $\parallel^l$ geometry), while the excited states (particularly 2p) are mostly suppressed. Further, the LA ground state MTCS falls off much slowly than the excited states through out the energy range ( 5-50 eV) for unequal energy sharing $(E_1 > E_2)$. These findings might have important implications for future $\overline{H}$ studies.

The single photon cross sections are significantly suppressed w.r.t. the FF indicating the importance of the multiphoton effects.

The $\parallel^l$ cross sections are much larger than the $\perp^r$ ones at low incident energies (for the spherical 1s & 2s states) while at higher energies the $\perp^r$ geometry overtakes. On the contrary, for the dipole allowed 2p state, the situation is just the reverse.

The $\overline{H}$ formation cross section is found to be quite sensitive w.r.t. the azimuthal angle $\phi$ for the $\perp^r$ geometry; a higher value of $\phi$ could be suggested for larger production.

For a more efficient formation of $\overline{H}$ for a wider energy range, the unequal distribution of energy ( $E_1 > E_2$ ) is preferred in both the LA and the FF cases.

The excited state dominance (over the 1s) in the FF TBR (for $\parallel^l$ geometry) at extreme low energies (~0-5 eV) is maintained in the laser assisted MTCS. However with increasing $E_1$, the 1s MTCS starts dominating over the 2s, 2p states.

For a greater yield of the ground state $\overline{H}$ than the excited ones through out the energy ( $E_1$ ) range, a higher value of $E_2$ using $\perp^r$ geometry could be suggested.

Finally, the present model is not supposed to be suitable for the experimental extreme low energy (~ meV) regime and a more sophisticated theory is needed, although the present approach is likely to be a pre-cursor to more difficult studies.

**Figure Captions:**

**Figure 1**: Multiphoton differential cross section MDCS ( in $a_0^2 sr^{-1}$ units ) vs scattering angle $\theta_f$ for laser assisted $\overline{H}$ formation via TBR process (1) for the active $e^+$ energy $E_1 = 10$ eV, the spectator $e^+$ energy $E_2 = 7$ eV in 1s, 2s, 2p states for $\parallel^l$ geometry ($\vec{\varepsilon}_0 \parallel \vec{k}_i$) and $\perp^r$ geometry ($\vec{\varepsilon}_0 \perp \vec{k}_i$). $\parallel^l$: solid line : 1s state, dashed line: 2s state and dotted line : 2p state; $\perp^r$ : solid line with squares: 1s state, solid line with triangles : 2s state , solid line with circles: 2p state. Laser parameters used: field strength $\varepsilon_0 = 5.0 \times 10^9$ V/m, photon energy $\hbar\omega = 1.17$ eV.

**Figure 2 (a)** Multiphoton total cross sections MTCS ( in units of $\pi a_0^2$ ) vs $E_1$ at a fixed $E_2 = 7$ eV for $\overline{H}$ formation in the ground (1s) state. Solid curve: $\vec{\varepsilon}_0 \parallel \vec{k}_i$; Dashed curve: $\vec{\varepsilon}_0 \perp \vec{k}_i$. Dotted curve: FF .**(b)** Same MTCS ( in units of $\pi a_0^2$) as in fig.2a but for the excited 2s state. **(c)** Same MTCS ( in units of $\pi a_0^2$) as in fig.2a but for the excited 2p state; laser parameters remaining the same.

**Figure 3 (a)** MTCS ( in units of $\pi a_0^2$) vs $E_1$ at a fixed $E_2 = 25$ eV for $\overline{H}$ formation in ground (1s) state and excited states (2s and 2p) for $\parallel^l$ geometry. Solid curve: 1s , dashed curve: 2s , dotted curve : 2p . Inset: corresponding FF TCS **(b)** Same MTCS as in fig.3a but for $\perp^r$ geometry; laser parameters remaining the same.

**Figure 4 (a)** Single photon total cross sections ( in units of $\pi a_0^2$) vs the number of photons emitted or absorbed in the ground state for $E_1$=10 eV, $E_2$ =7 eV; filled square: results for $\vec{\varepsilon}_0 \parallel \vec{k}_i$, open square : results for $\vec{\varepsilon}_0 \perp \vec{k}_i$ **(b)** Same TCS ( in units of $\pi a_0^2$) as in fig.4a but for the 2s state, **(c)** Same TCS ( in units of $\pi a_0^2$) as in fig. 4a but for the 2p state.

**Figure 5** Variation of the MTCS (in units of $\pi a_0^2$) with respect to the azimuthal angle $\phi$ for $\perp^r$ geometry ; other laser parameters remaining the same as in previous figures. Solid curve : 1s state, dashed curve : 2s state, dotted curve : 2p state.

**Figure 6** Ground state MTCS (in units of $\pi a_0^2$) vs the laser photon energy ( in eV) for $E_1 = 10$ eV, $E_2 = 7$ eV; other laser parameters remaining the same as in previous figures. Solid curve: $\parallel^l$ geometry; dashed curve: $\perp^r$ geometry.

**Figure 7** Ground state MTCS ( in units of $\pi a_0^2$) vs the laser field strength in a.u. for $E_1 = 10$, $E_2 = 7$ eV. Solid curve: $\vec{\varepsilon}_0 \parallel \vec{k}_i$; dashed curve: $\vec{\varepsilon}_0 \perp \vec{k}_i$; dotted curve : FF.

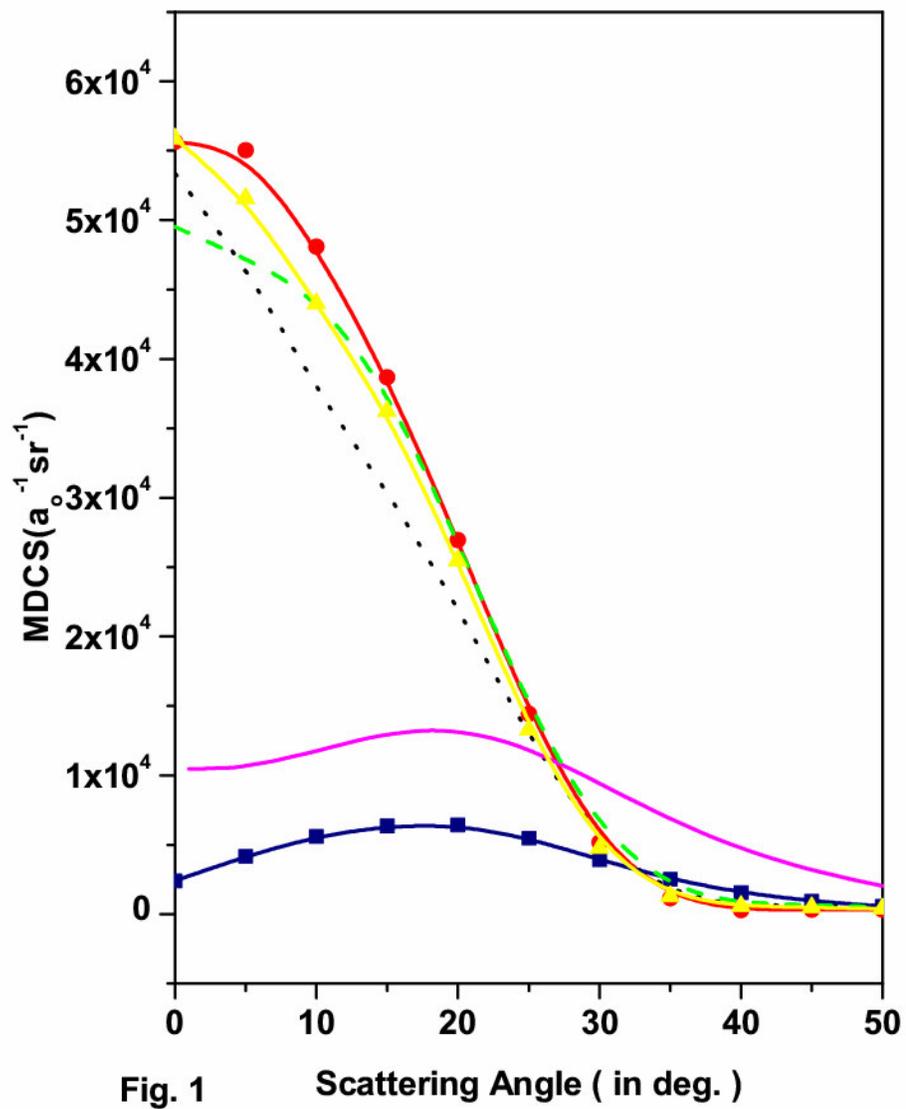

Fig. 1 Scattering Angle ( in deg. )

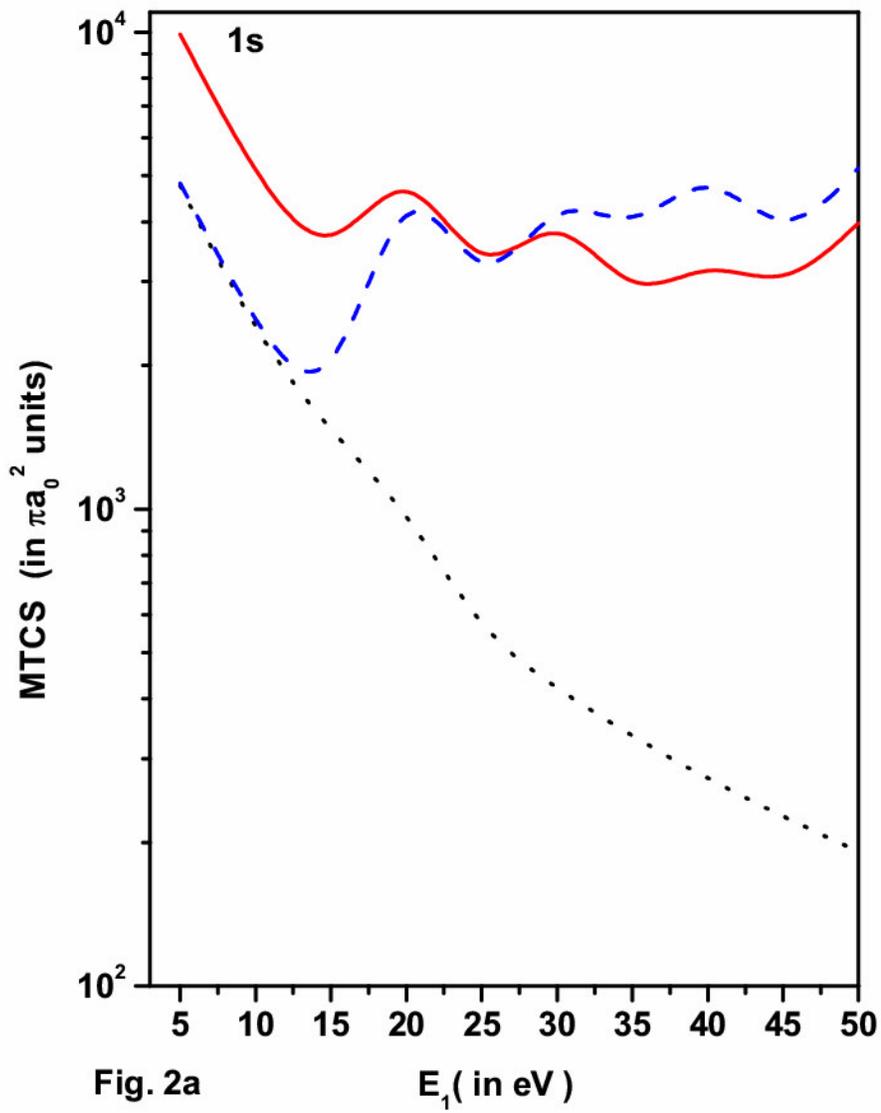

Fig. 2a

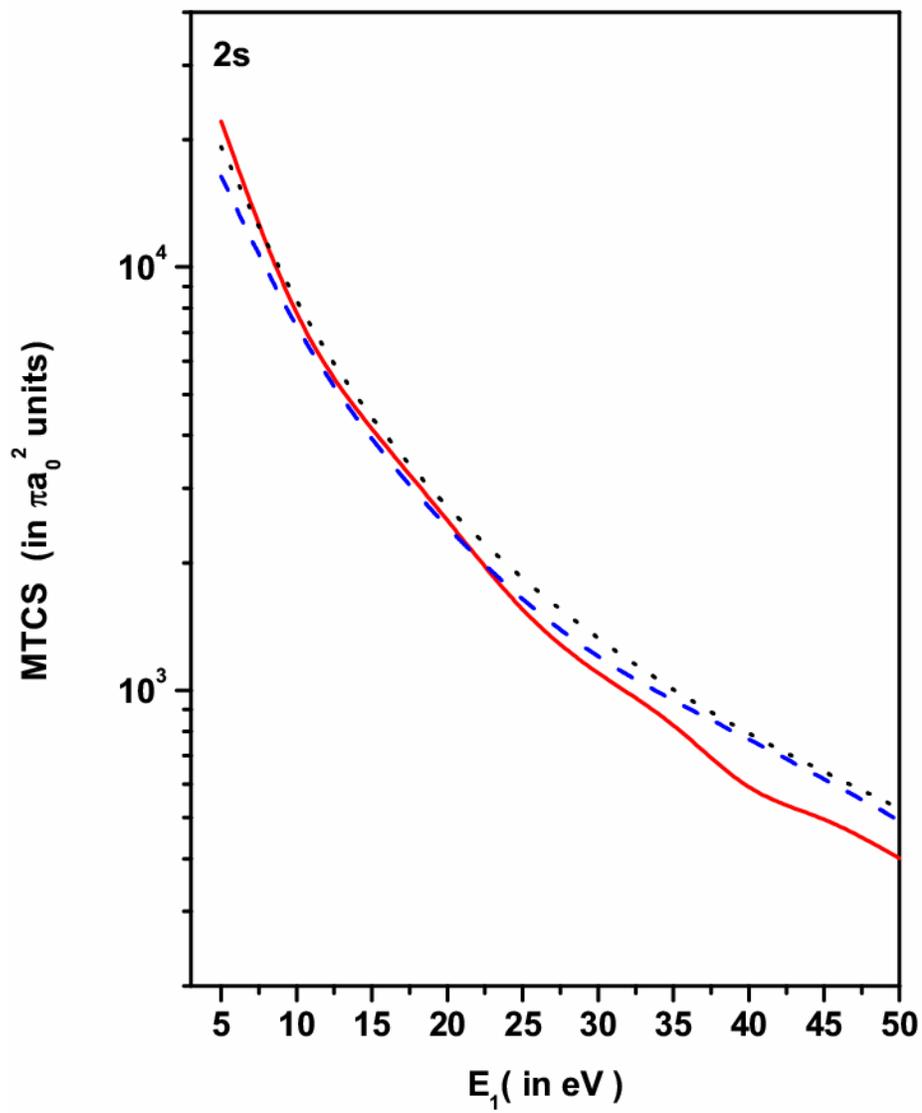

Fig. 2b

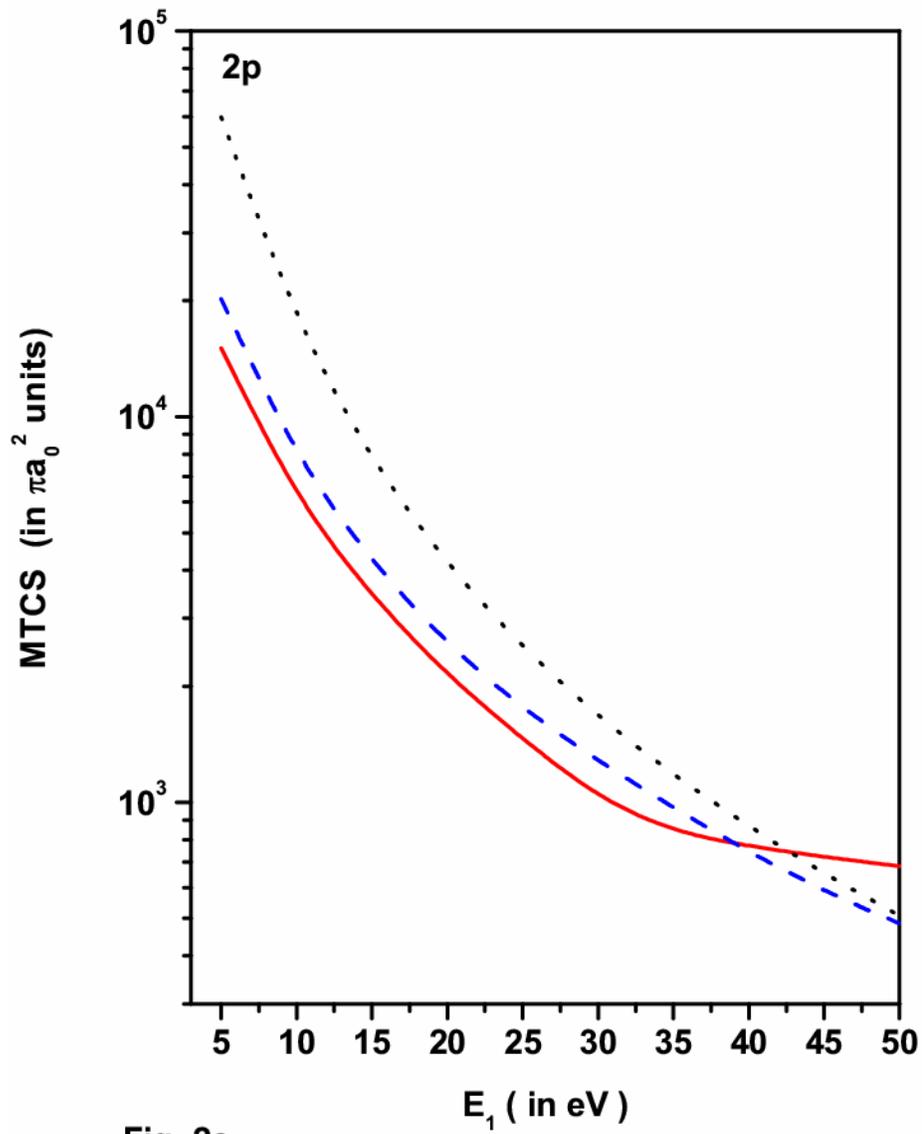

Fig. 2c

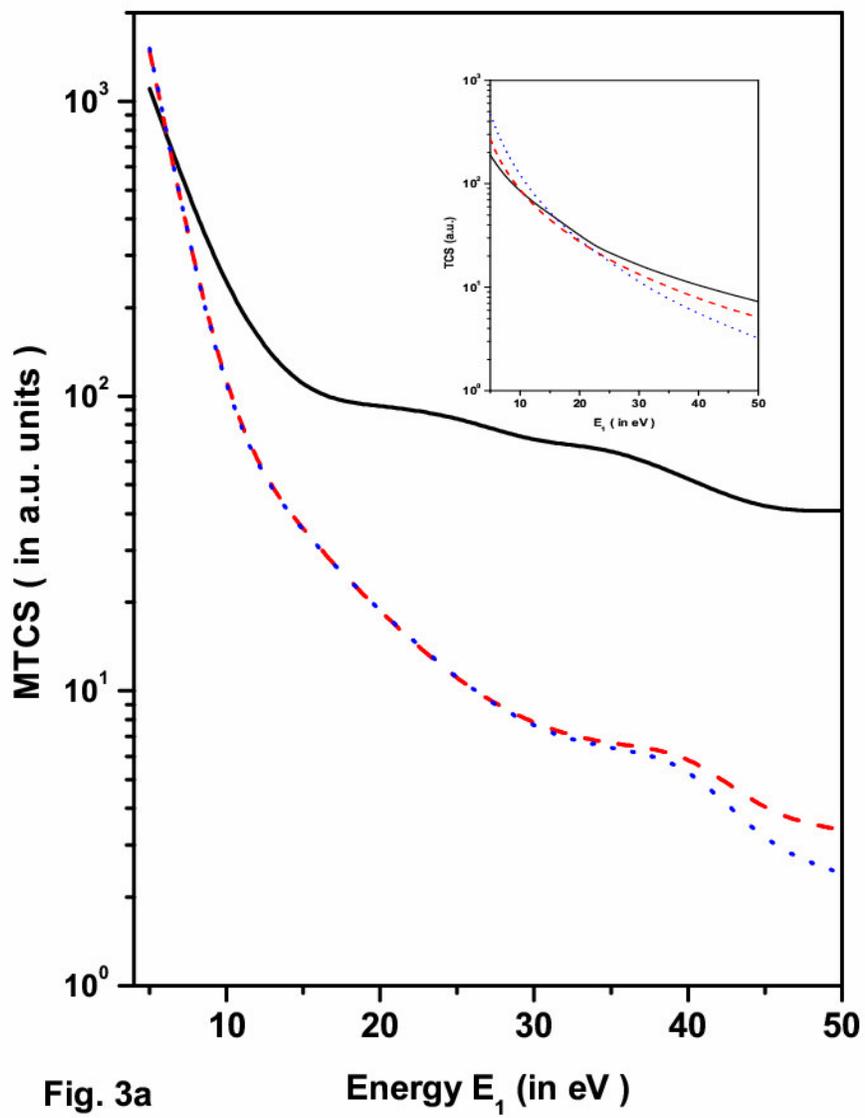

Fig. 3a

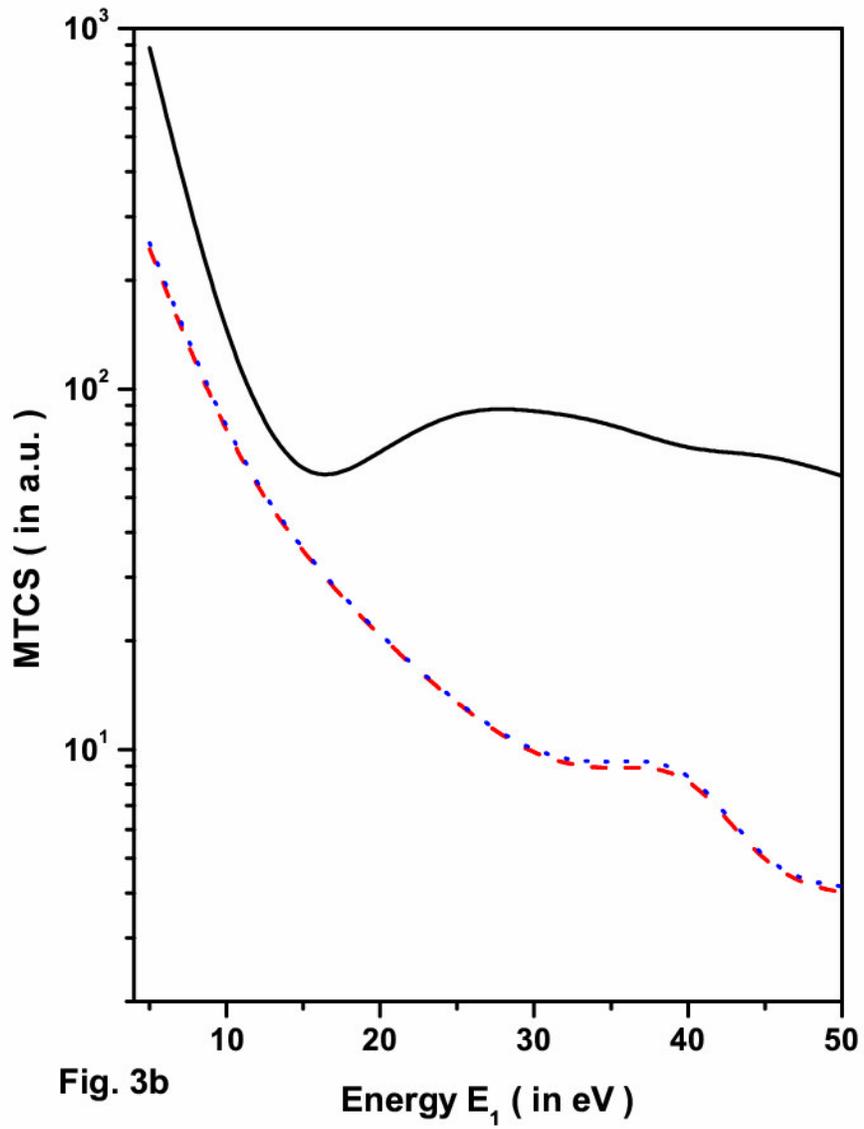

Fig. 3b

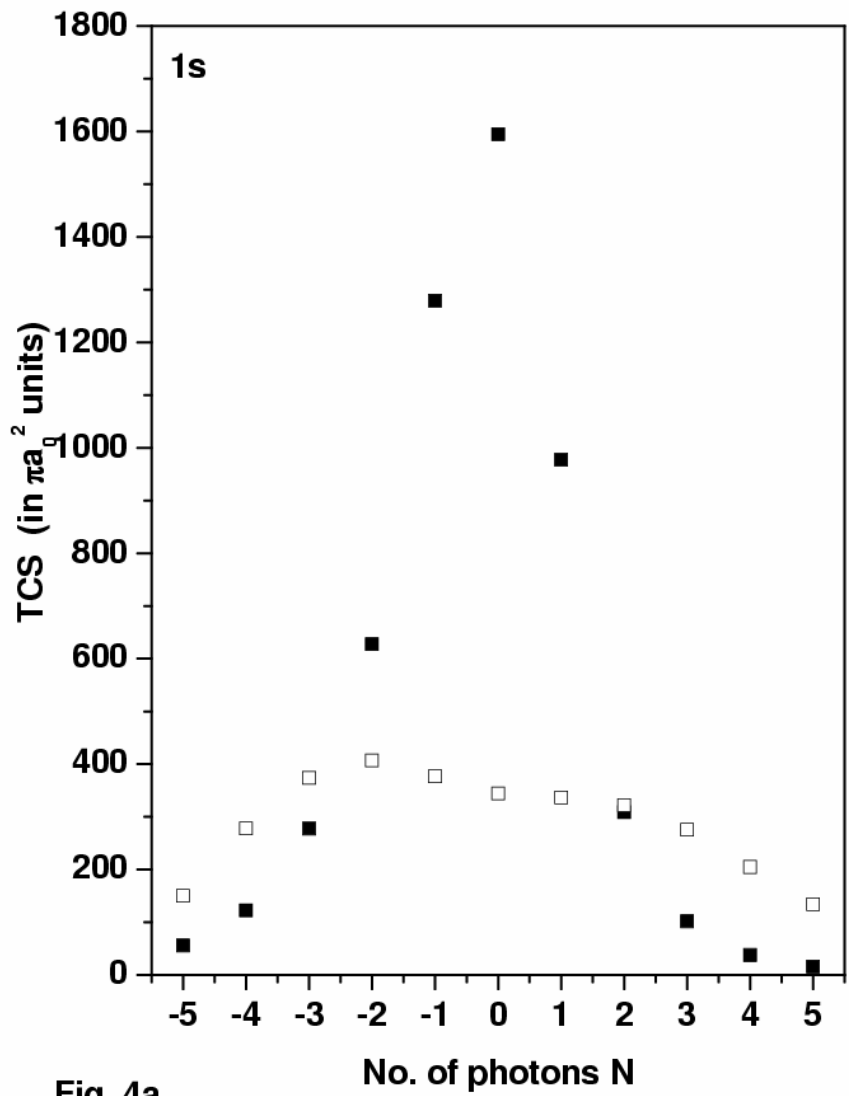

Fig. 4a

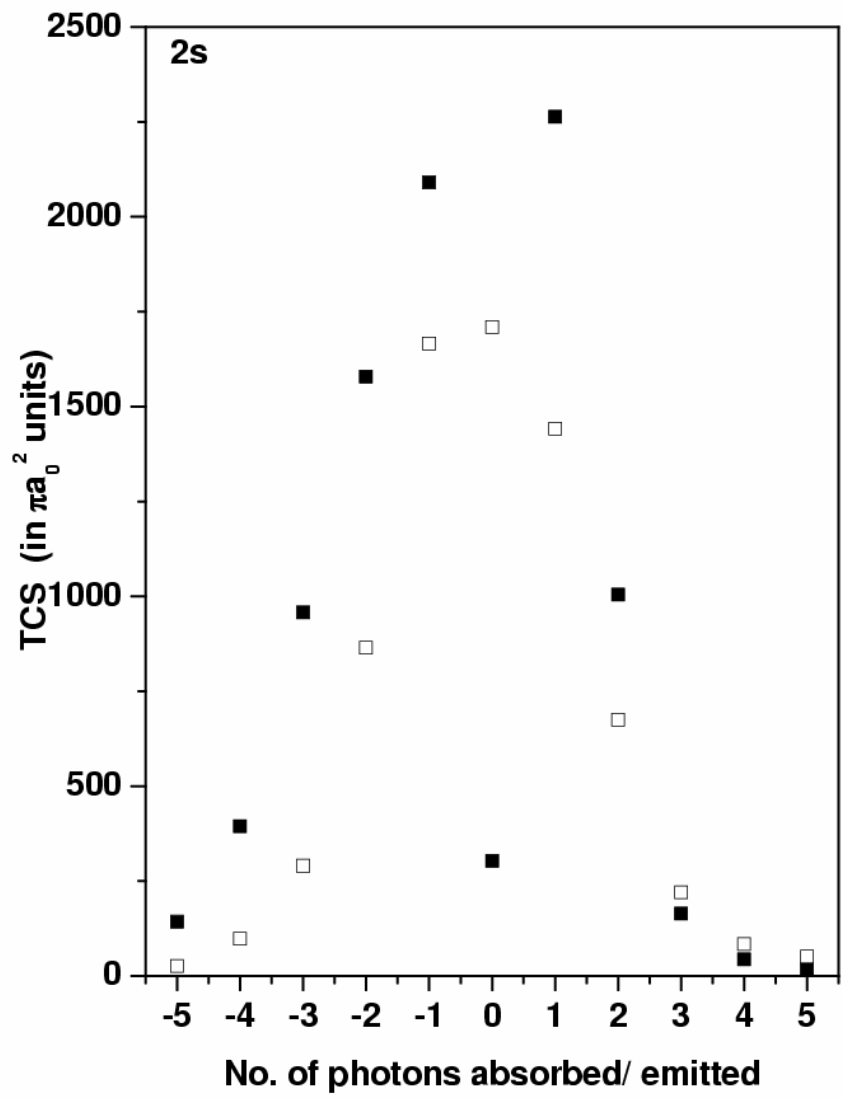

Fig. 4b

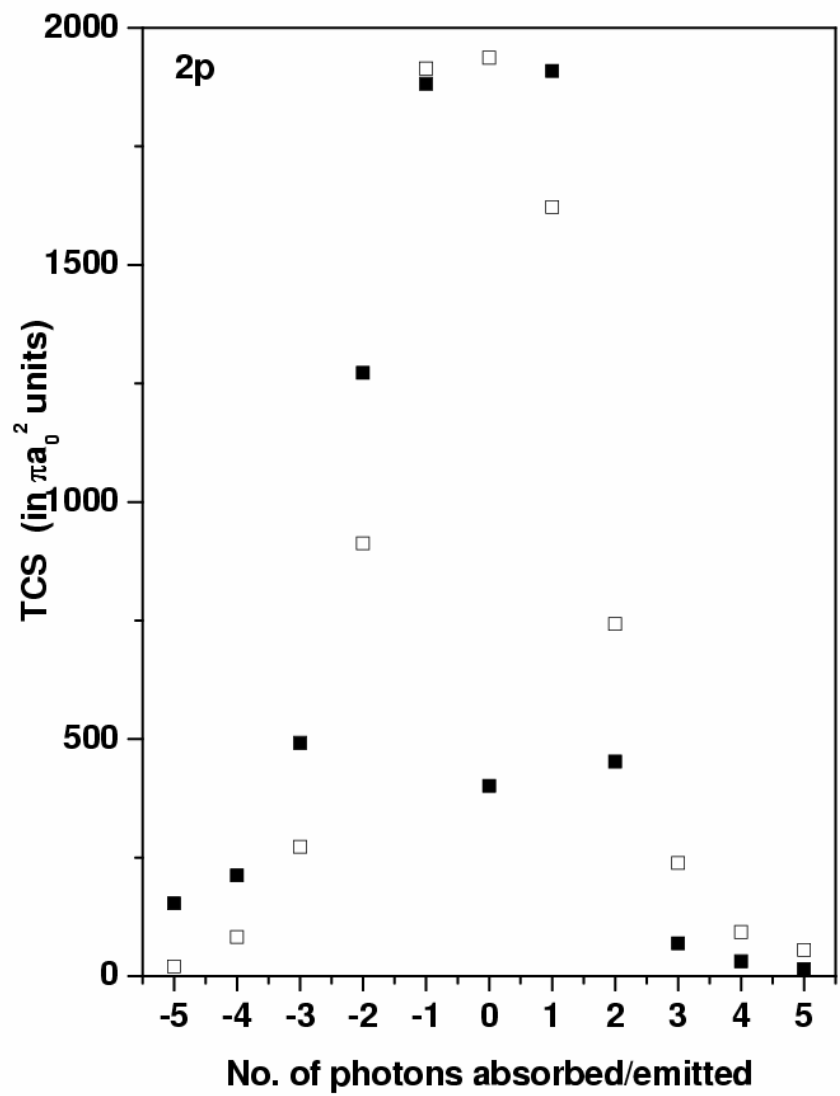

Fig. 4c

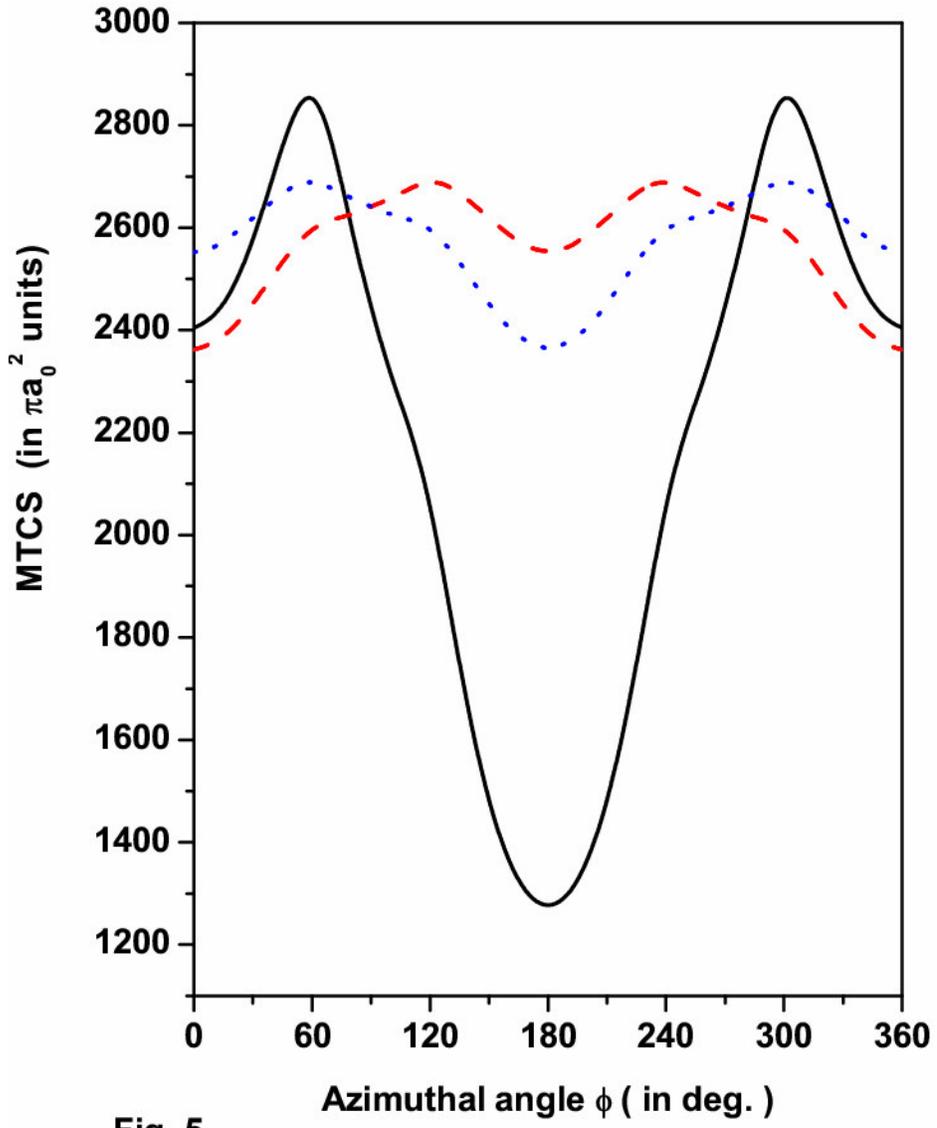

Fig. 5

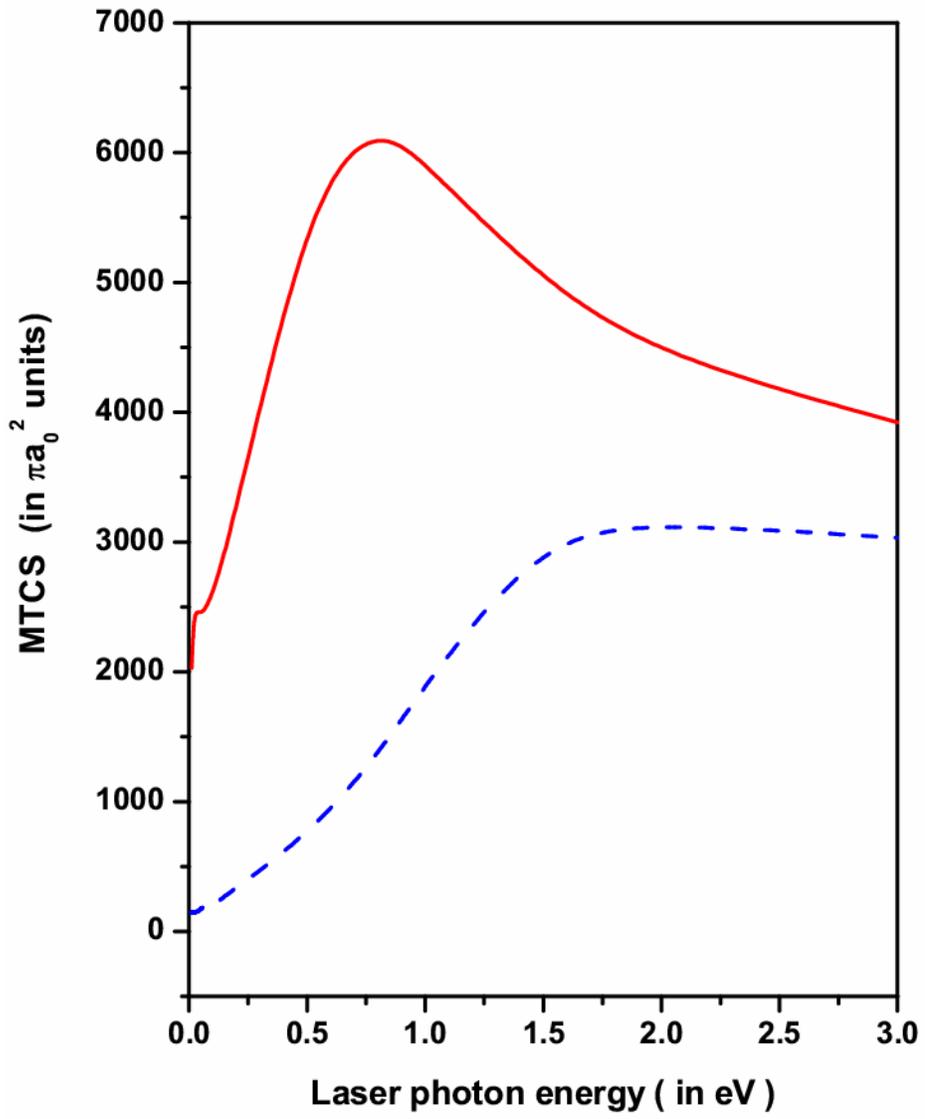

**Fig. 6**

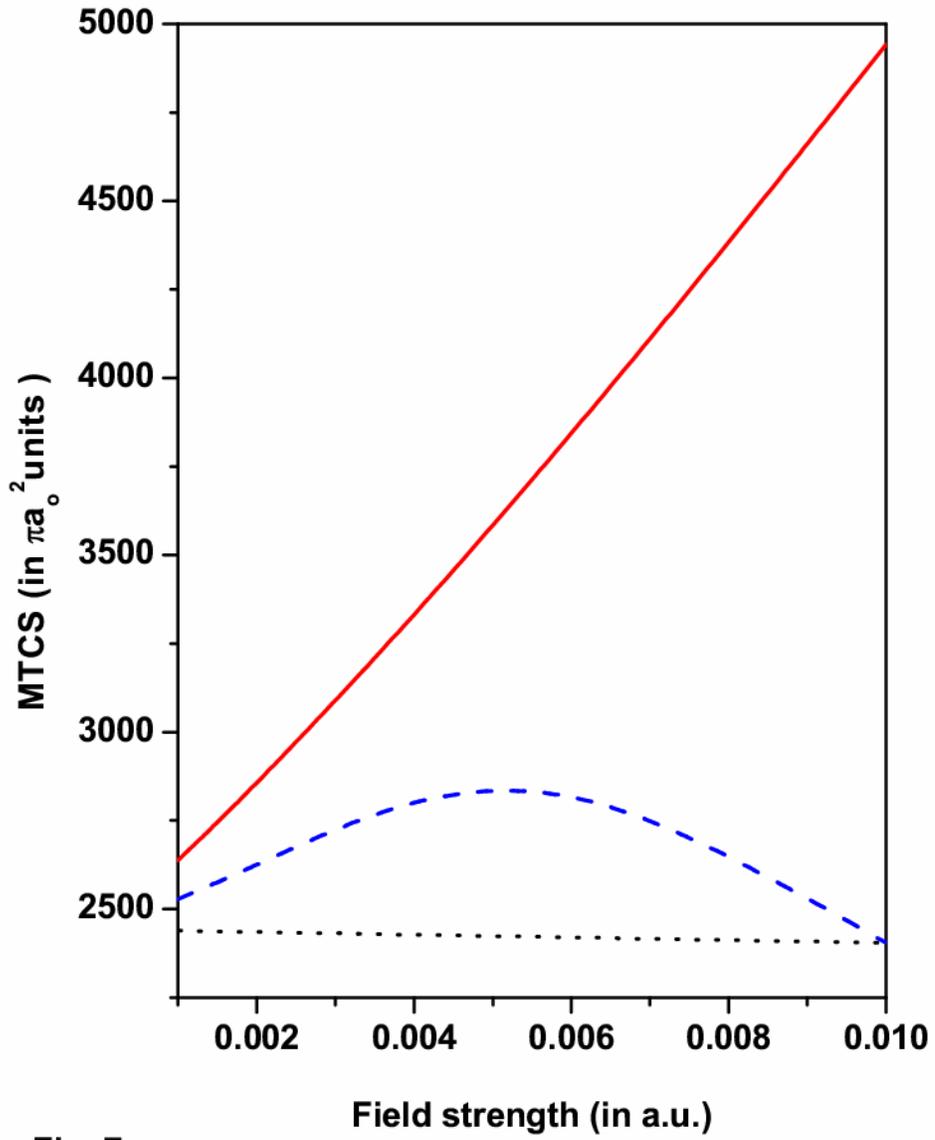

Fig. 7